\documentclass[]{revtex4-1}
\usepackage{amsmath}    
\usepackage{graphicx}   
\usepackage{verbatim}   
\usepackage{color}      
\usepackage{subfigure}  
\usepackage{hyperref}   
\usepackage{lineno} 
\usepackage{bm} 
\usepackage{multirow}
\usepackage{csquotes}
\usepackage{appendix}
\newcommand{\classoption}[1]{\texttt{#1}}

\expandafter\ifx\csname package@font\endcsname\relax\else
\expandafter\expandafter
\expandafter\usepackage
\expandafter\expandafter
\expandafter{\csname package@font\endcsname}%
\fi
\DeclareRobustCommand\substyle{\name@idx{document substyle}}%
\DeclareRobustCommand\classoption{\name@idx{document class option}}%
\DeclareRobustCommand\classname{\name@idx{document class}}%
\def\name@idx#1#2{
	{\ttfamily#2}%
	\index{#2\space#1=\string\ttt{#2}\space#1}\index{#1>#2=\string\ttt{#2}}%
}

\begin{document}

	\pagenumbering{arabic}
	\title{Open Charm production in proton-proton collisions at $\sqrt{s}$ = 13.6 TeV with Pythia event generator}
\author{ Randhir Singh$^{1}$\footnote{E-mail: randhir.j5246@cgcuniversity.in}}

\affiliation{Department of Applied Sciences, School of Engineering and Technology,
	CGC University, Mohali-140307, Punjab$^{1}$}

	\begin{abstract}
 Charm and beauty are heavy quarks with observed masses of 1.28 GeV/$\textit{c}^2$ and 4.18 GeV/$\textit{c}^2$ respectively. They are produced in initial hard scattering processes. Due to their small formation time ($\Delta t \sim0.1 fm/\textit{c}$) as compared to the formation time of QGP ($\Delta t \sim0.3 fm/\textit{c}$) at the LHC, they experience all the stages occuring during the time evolution of the hot and dense medium produced in heavy-ion collisions. Therefore, the measurement of open charm and beauty production allows probing QGP properties and investigating the color charge and mass dependence of the parton in-medium energy loss. Moreover, due to their large masses ($m_c , m_b \gg \Lambda_{QCD}$ ) their pp production cross-sections are calculable within the domain of perturbative QCD constituting an excellent test of pQCD calculations. The aim of this study is to understand the processes involved in the production of charm quarks through the productions of D$^0$, D$^+_s$ and $\Lambda_c^+$ hadrons. Further to investigate the possibility of hadronization of the charm quarks, ratios like $\Lambda_c^+$/$D^0$ and $D_s^+$/$D^0$ are also measured. For the current analysis, the events are generated by using PYTHIA 8 for pp collisions at $\sqrt{s}$ = 13.6 TeV. PYTHIA 8 has proved to be quite successful in explaining the heavy-flavor particle production at the  LHC energies.\\
 
	\end{abstract}
	\maketitle

\section{Introduction}
\label{section1}
The hot, dense and strongly interacting state of matter, called as simply quark-gluon plasma (QGP) is created in heavy-ion collision experiment \cite{1}. The properties of this hot and dense system cannot be measured directly. Instead the new particles produced from this system carry signals of QGP, and its properties which can be directly and indirectly measured. One of the very important analysis tools to understand the evolution dynamics of this system is provided by the heavy quark flavours. The Heavy flavours like charm and bottom quark play a significant role in the study of QGP properties\cite{2}. Due to their large masses, they are produced in the hard partonic scattering processes occuring at very early phases of ultra-relativistic heavy-ion collisions. These heavy quarks have longer mean life-time than the QGP. Therefore, they experience all stages occuring in the  formation and evolution of the QGP system. A study of production mechanisms of heavy quarks is helpful to explore the pQCD because their mass(m$_Q$) provide a scale at which the strong interaction coupling constant($\alpha_{s}$) is generally evaluated \cite{3}. As m$_Q$ $>>$ $\Lambda_{QCD}$, the production mechanism should be calculated under pQCD regime. The difference in mass of these quarks is useful to probe different QCD regions at different Q$^2$ values.\\
The open heavy-flavour production measurement in pp collisions as a function of the charged-particle multiplicity could provide insight into the processes occurring in the collision at the partonic level and the interplay between the hard and soft mechanisms in particle production. These aspects are expected to have dependence on the energy and on the impact parameter of the pp collision\cite{4,5,6}. In particular, pp collisions in which hard parton-parton scattering takesplace are predicted to be more central in nature than minimum-bias events \cite{7,8}. A 20$\%$ higher average charged-particle multiplicity was observed by NA27 Collaboration in 1988 in the events in which  open charm production takesplace in comparison the events without charm production \cite{9}. The softening of the momentum spectra of hadrons produced in association with charm was also observed. This result was interpreted as a consequence of the more central nature of collisions leading to charm production.\\
At LHC energies, there are two more additional factors which contribute to charm production and its relation to multiplicity of the events. The first of these is the larger amount of gluon radiation which is associated to the short distance production processes at larger energies and particle transverse momenta. The second effect is the contribution of Multiple-Parton Interactions (MPI) \cite{10,11,12}, i.e. several hard partonic interactions occurring in a single pp collision. The pQCD-based models of MPIs describe high multiplicity pp collisions. One such model is PYTHIA 8 \cite{13}, which is a Monte Carlo event generator. The ALICE Collaboration performed the analysis on minijet production and the results indicates that high multiplicities in pp collisions are reached through a high number of MPIs \cite{14}. In an another analysis performed in the heavy-flavour sector by the LHCb, the production of double charm production ($D^0 + X$, $J/\psi + X$ and $J/\psi +J/\psi$ where $X= D^0$, $D^+$, $D^+_s$ and $\Lambda_c$) in pp collisions at the LHC has been measured and the results suggest MPIs also play a role at the hard momentum scale relevant for $c\overline{c}$ production \cite{15,16,16a}.\\

In this analysis, we have studed the production of open charm flavour hadrons ($D^0$, $D^+_s$ and $\Lambda_c^+$) through the $p_T$ spectra and the self normalised yield as a function of relative charged particle density. In addition to that the ratio of strange to non-strange mesons is calculated. Also baryon to meson ratio has been calculated to investigate the possibility of hadronization of the charm quarks. The analysis is performed with the \emph{pp} data generated by PYTHIA 8 Monte Carlo event generator at $\sqrt{s}$ = 13 TeV. A similar analysis was done at $\sqrt{s}$ = 13.6 TeV using pp collision data from ALICE detector at the LHC where production yields of D$^0$, D$^+_s$, and $\Lambda_c^+$ \cite{16b} have been measured. Also the ratios like $\Lambda_c^+$/$D^0$ and $D_s^+$/$D^0$ have been measured.\\

The paper is organised as follows. We begin with a brief motivation for the study in Section \ref{section1}. In Section \ref{section2}, the detailed analysis methodology along with brief description of PYTHIA 8 are given. Section \ref{section3} discusses the results and finally they are summarized in Section \ref{section4}. 

\section{EVENT GENERATION AND ANALYSIS	METHODOLOGY}
\label{section2}
\begin{figure}[h!]
	\centering     
	\subfigure[] {\label{fig:a}\includegraphics[scale=0.15]{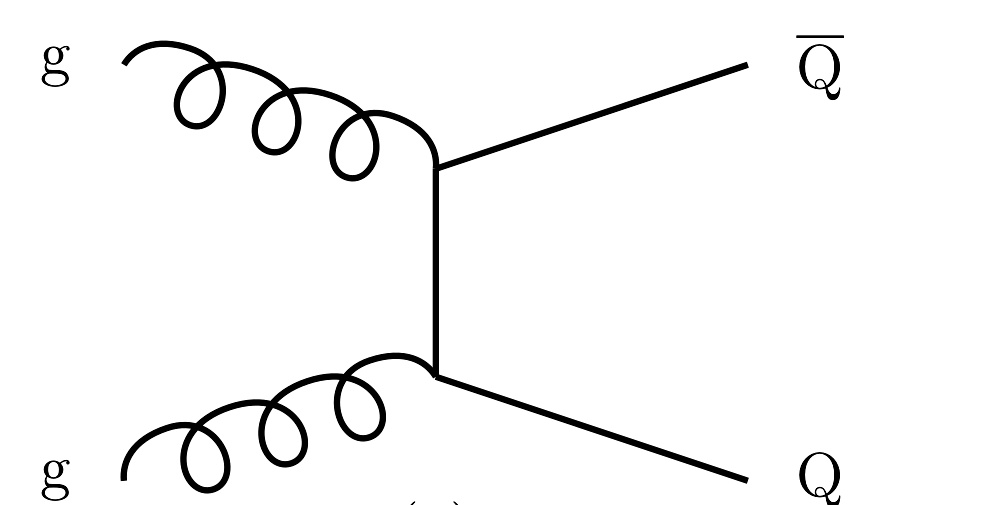}}
	\subfigure[] {\label{fig:b}\includegraphics[scale=0.15]{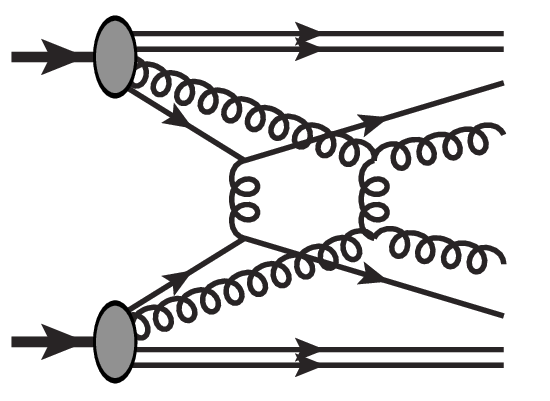}}
	\subfigure[] {\label{fig:c}\includegraphics[scale=0.15]{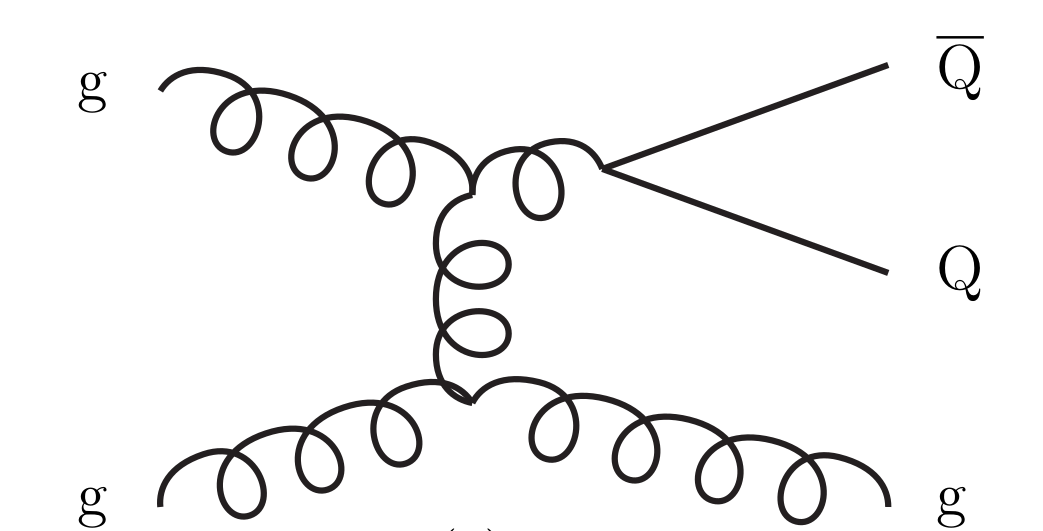}}
	\subfigure[] {\label{fig:d}\includegraphics[scale=0.18]{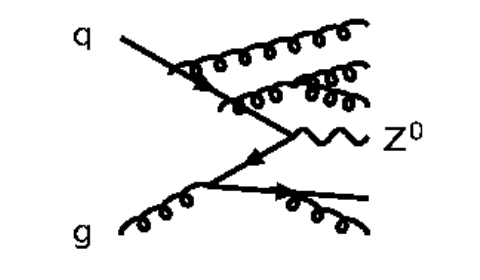}}
	\subfigure[] {\label{fig:e}\includegraphics[scale=0.18]{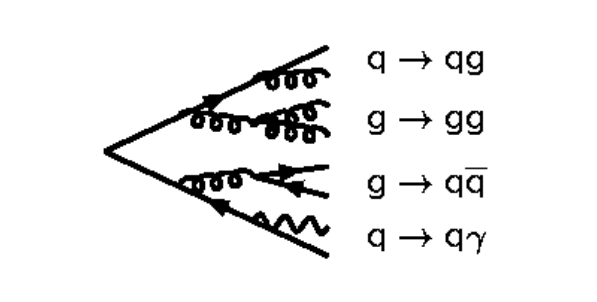}}
	\caption{Examples of heavy-flavour production diagrams. (a) Gluon fusion. (b) Multi-partonic interaction (c) gluon splitting (d) Initial state radiation (e) Final state radiation}
	\label{processes}
\end{figure}

In real experiments accelerators are used to colliide the particles at relatrivistic speeds. As a result of which many new particles are produced. But in Monte Carlo simulation different event generators are used to simulate events.\\
Pythia is one of the general-purpose Monte Carlo event generator. It is a software package for the simulation of high energy physics events.  It simulates ultra-relativistic collisions between particles like like electron-electron, electron-positron, proton-proton and proton-antiproton. It simulates physical phenomenon like MPI, Hard and soft events, initial-state radiations(ISR) and final-state radiations(FSR), parton shower, fragmentation and decay processes.\\
Pythia contains a richer selection of Hard processes(around 300 hard processes). If we categorise them on the basis of number of final state particles produced, Pythia is well equipped for 2 $\rightarrow$ 1 and 2$\rightarrow$ 2 processes and for more number of particles in final state. On the basis of physics scenario, it implements processes like Hard QCD processes (i.e. $q\overline{q}$ $\rightarrow$ $q\overline{q}$, $q\overline{q}$ $\rightarrow$ $gg$, $gg$ $\rightarrow$ $q\overline{q}$, $gg$ $\rightarrow$ $gg$ etc.), Soft QCD processes (Elastic ($AB$ $\rightarrow$ $AB$), Single diffractive ($AB$ $\rightarrow$ $XB$ or $AB$ $\rightarrow$ $AX$), Double diffractive ($AB$ $\rightarrow$ $X$), Non diffractive, Heavy flavour production (i.e. $gg$ $\rightarrow$ b$\bar{b}$($c\overline{c}$) and $q\overline{q}$ $\rightarrow$ b$\bar{b}$($c\overline{c}$)), Prompt-photon production (qg $\rightarrow$ q$\gamma$), Deep-inelastic scattering (ql $\rightarrow$ ql), Standard Model Higgs boson production and many more. Since the gluons produced in scattering processes leads to heavy-flavour production, in Pythia partonic interactions are described via initial-state radiations and final-state radiations. There are some multi-partonic interactions which occur in addition to Hard processes where generally more than one parton interaction occur. Transverse-momentum ordered showers form the basis of both initial-state radiations and final-state radiations and for multi-partonic interactions, matrix element formulation is used  in case of 2 $\rightarrow$ 2 processes. Also initial-state radiations and final-state radiations are combined in multi-partonic interactions using decreasing p$_T$ order.\\ 
The production of Heavy-flavour in PYTHIA 8 proceeds via four main mechanisms: (i) The first one (hardest) are hard processes, in which the initial c/b quarks originate from the first 2$ \rightarrow$ 2 hard process, mostly by gluon fusion ($gg\rightarrow c\overline{c}$ or involving a c/b sea-quark (e.g. $cu \rightarrow cu$) (ii) Second one are the hard process in MPI, which are produced via the same mechanisms as the first hard process but in consecutive interactions. (iii) Third one proceeds by a proces called gluon splitting from hard process in which each produced gluon(originates from a hard process, either the first one or a subsequent process (in MPI) has a probability to split into a $c\overline{c}$ or $b\overline{b}$ pair contributing to heavy-flavour production. (iv) When the initial gluon is produced from initial or final state radiation, we refer to this process as ISR/FSR. It should be noted that the biggest contribution to hard processes in PYTHIA 8 comes from c sea-quarks and not from gluon fusion\cite{17}. The different processes are shown in fig.\ref{processes}.\\
In Pythia, MPI and CR are two default settings and both have significant affect on charged-particle multiplicity. The CR and MPI effects are strongly correlated. But multi-parton interaction is important to study heavy flavour production mechanisms. As decribed in hadronisation process, in Pythia heavy-quarks are produced in high p$_T$ transfering scattering processes and are indroduced via pQCD matrix element method. In Pythia8, the kinematics of the hard-process system is not reprocessed after ISR, thereby preserving the original (Born-level) kinematic configuration. Detailed explanation on PYTHIA 8 physics processes and their implementation can be found in Ref.\cite{18}.\\
The results reported in this paper are obtained by simulating inelastic events using hard QCD mode (HardQCD: hardccbar (hardbbbar)=on) with PYTHIA 8 event generator. This configuration includes the processes $q\overline{q}$ $\rightarrow$ $b\overline{b}$, $gg$ $\rightarrow$ $b\overline{b}$, $q\overline{q}$ $\rightarrow$ $c\overline{c}$ and $gg$ $\rightarrow$ $c\overline{c}$. In addition to that Color Reconnection tune, CR2 (called Gluon-move model)\cite{19} is used which considers Tune:pp = 14 and which include reconnection beyond leading colour. In the CR2 model only the gluons are allowed to participate in the reconnection process. For each gluon all the reconnections formed to all MPI systems are considered (in addition to the ones for softer MPIs), therefore in principle it affects more significantly the colour flow from the hard interaction. A cut of $p_T \geq$ 0.5 GeV/\textit{c} (using PhaseSpace:pTHatMinDiverge) is used to avoid the divergences of QCD processes in the limit $p_T$ $\rightarrow$ 0. Study of D$^0$, D$^+_s$, and $\Lambda_c^+$ production are done at the mid-rapidity ($|y| < 0.5$) and in the kinematic range [1,24] GeV/\textit{c}. D$^0$, D$^+_s$, and $\Lambda_c^+$ are selected by using their respective pdg values. This analysis is performed by generating 175 million pp events at $\sqrt{s}$= 13.6 TeV. The charged-particle multiplicity, N$_{ch}$ is measured at the mid-rapidity ($|\eta| < $1.0). The analysis has been performed in the four  multiplicity classes [1-15],[16-30],[31-60],[61-100]. The mean values for the charged particle density in each multiplicity bin are shown in the table.\ref{table2}.

\begin{table}
	\centering
	\begin{tabular}{|c|c| }
		\hline
		N$_{ch}$ & $<dN_{ch}$/$d\eta >$\\
		\hline
		[1-15] & 3.76 \\
		\hline
		[16-30] & 	11.72\\
		\hline
		[31-60] & 	23.14\\
		\hline
		[61-100] & 	39.88\\
		\hline 
	\end{tabular}
	\caption{Multiplicity classes and average charged-particle density.}
	\label{table2}
\end{table}

\section{Results and Discussion}
\label{section3}
\begin{figure}[h!]
	\centering     
	\subfigure[] {\label{fig:a}\includegraphics[width=3.0in,height=2.0in]{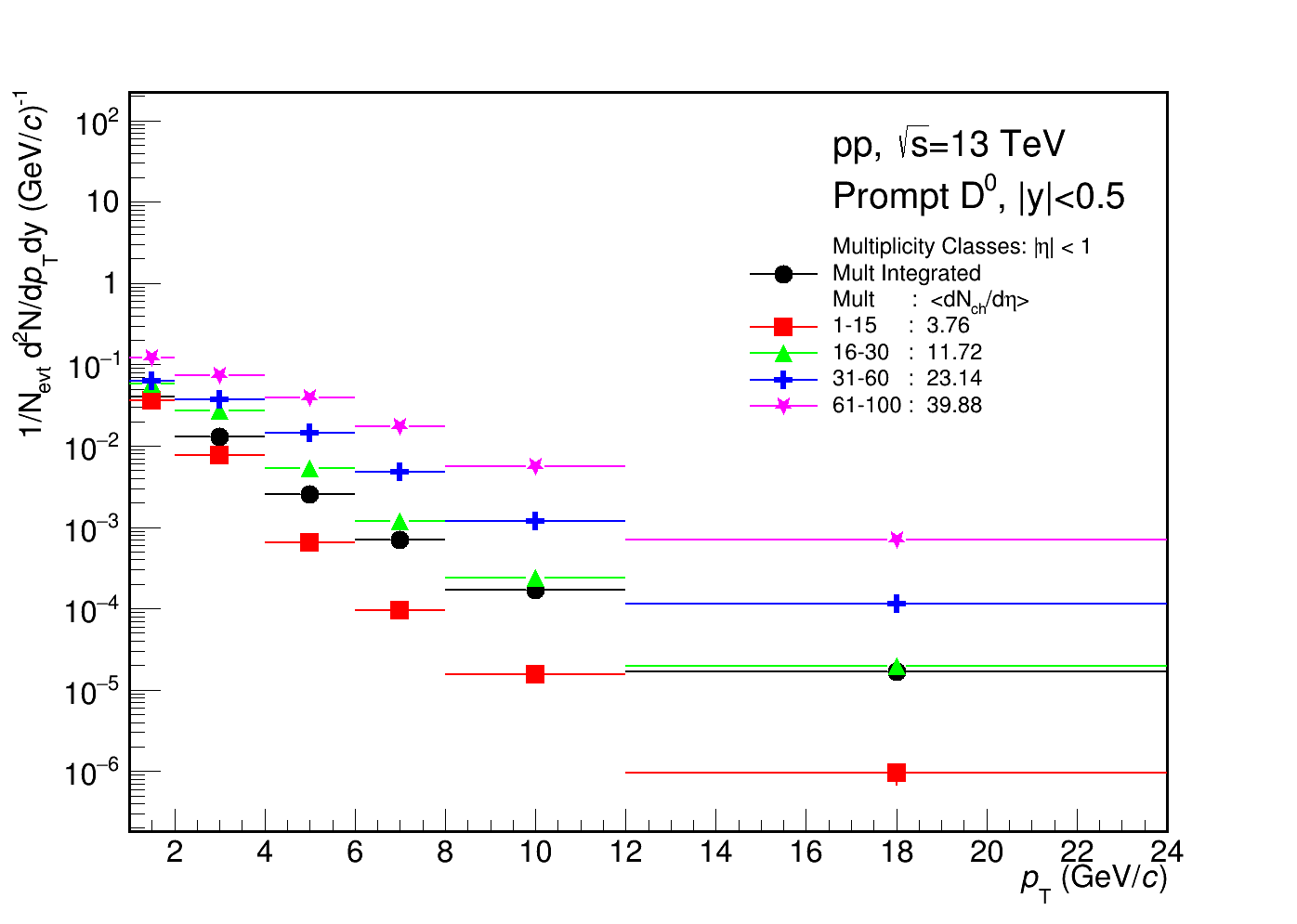}}
	\subfigure[] {\label{fig:b}\includegraphics[width=3.0in,height=2.0in]{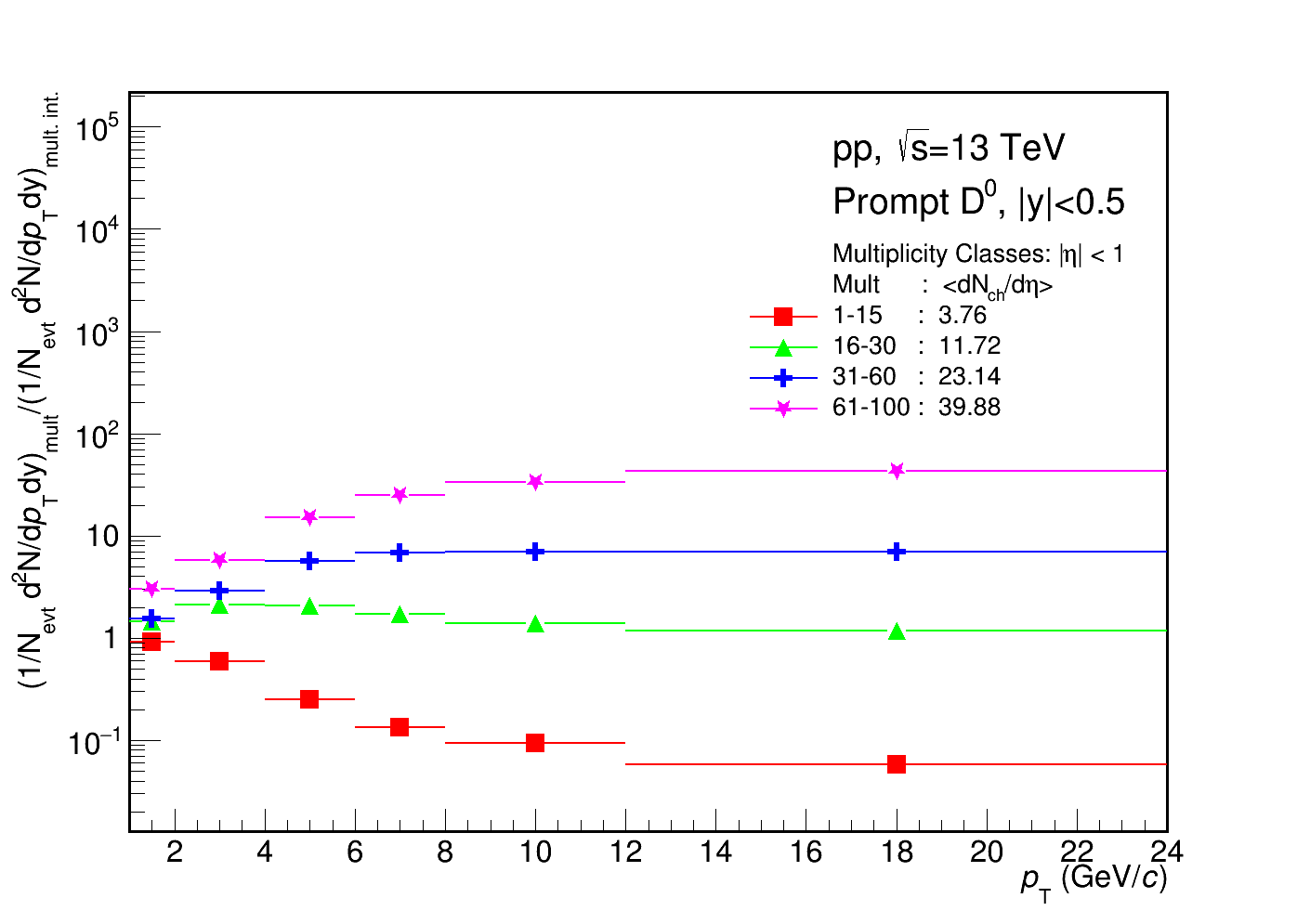}}
	\caption{(Left) Corrected spectra for D$^0$-meson in the integrated and different multiplicity ranges. (Right) Ratio of corrected spectra in different multiplicity bins with respect to the spectra in the integrated multiplicity bin}
	\label{figureDzero}
\end{figure}

\begin{figure}[h!]
	\centering     
	\subfigure[] {\label{fig:a}\includegraphics[width=3.0in,height=2.0in]{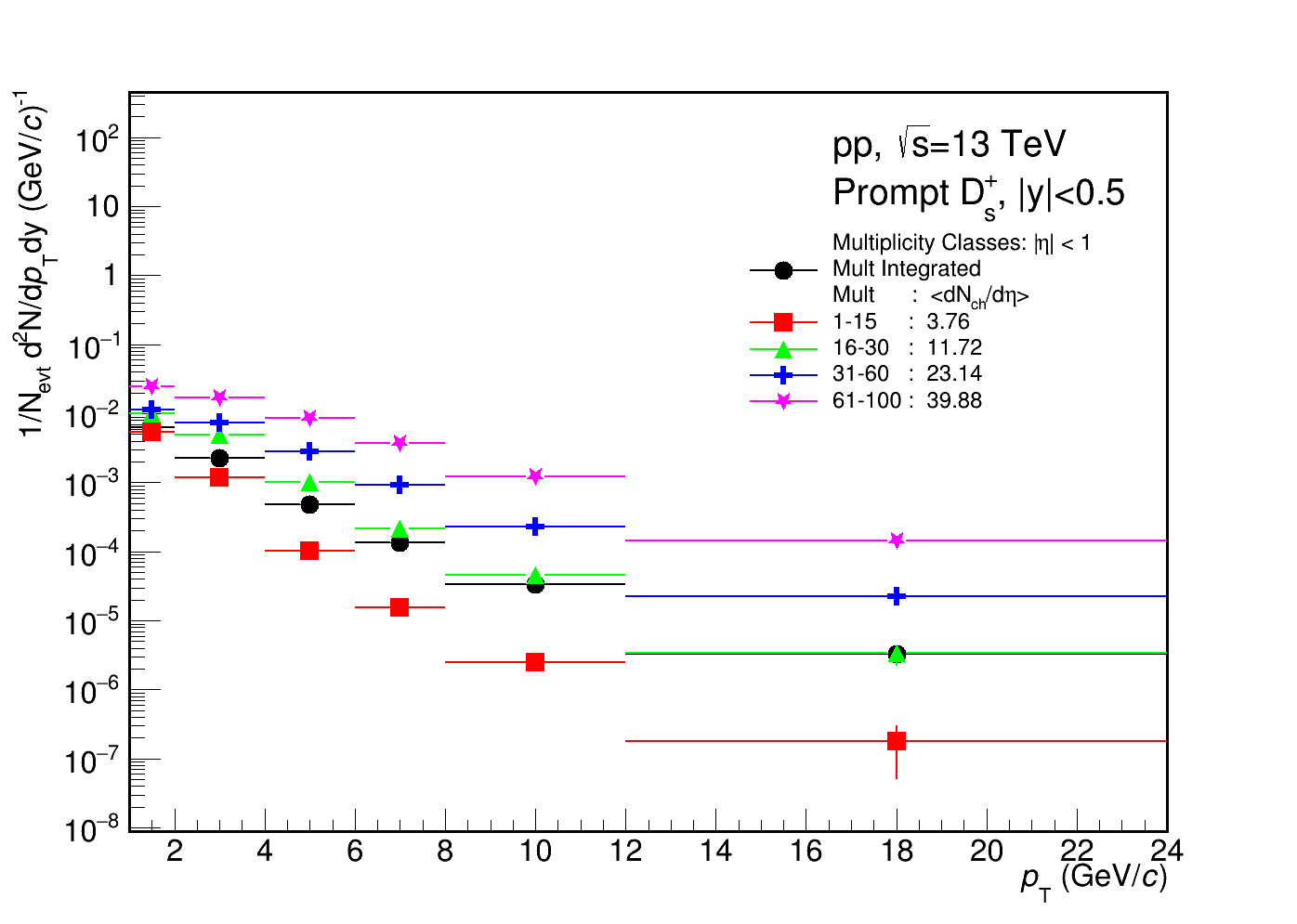}}
	\subfigure[] {\label{fig:b}\includegraphics[width=3.0in,height=2.0in]{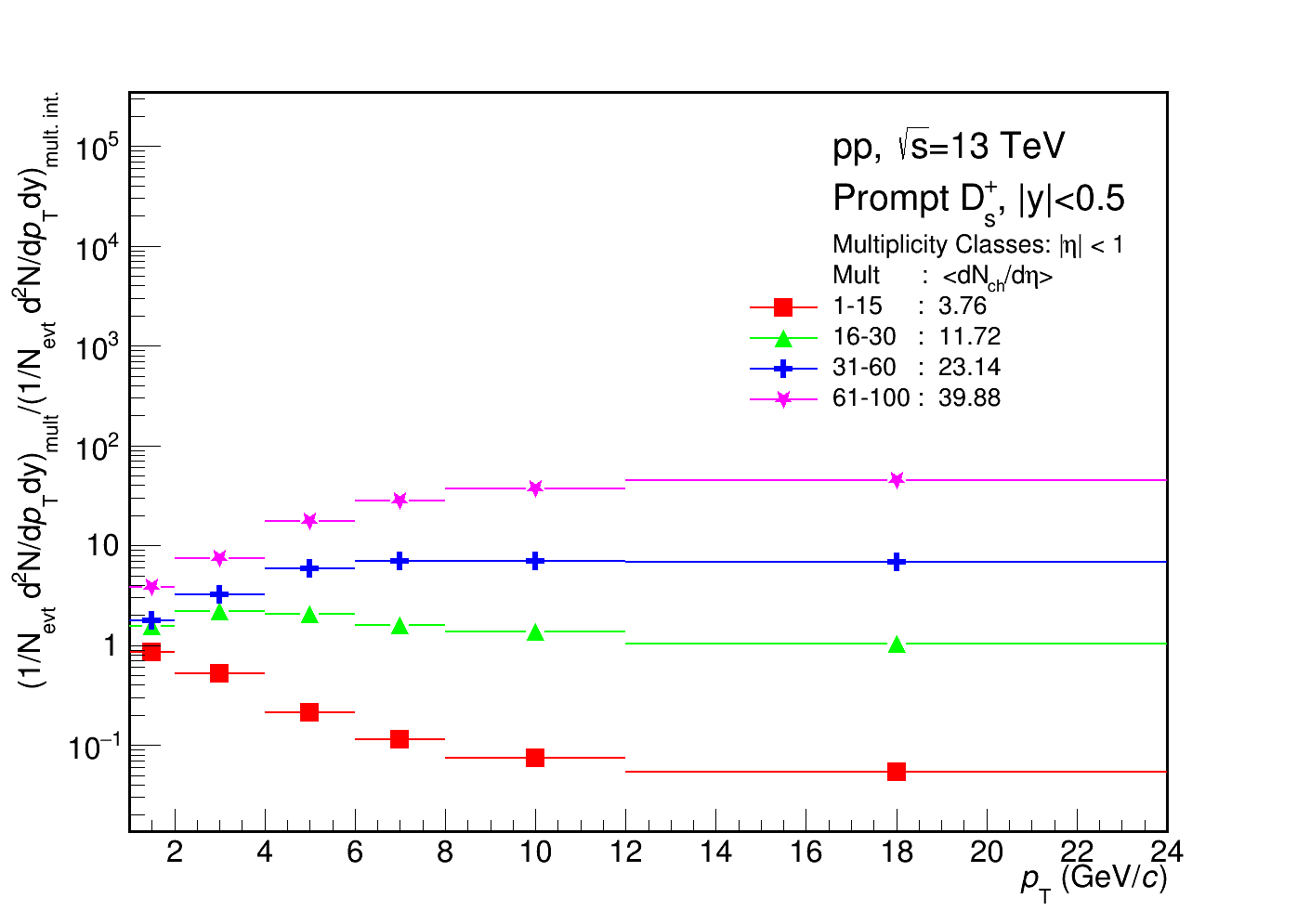}}
	\caption{(Left) Corrected spectra for D$^+_s$-meson in the integrated and different multiplicity ranges. (Right) Ratio of corrected spectra in different multiplicity bins with respect to the spectra in the integrated multiplicity bin}
	\label{figureDsplus}
\end{figure}

\begin{figure}[h!]
	\centering     
	\subfigure[] {\label{fig:a}\includegraphics[width=3.0in,height=2.0in]{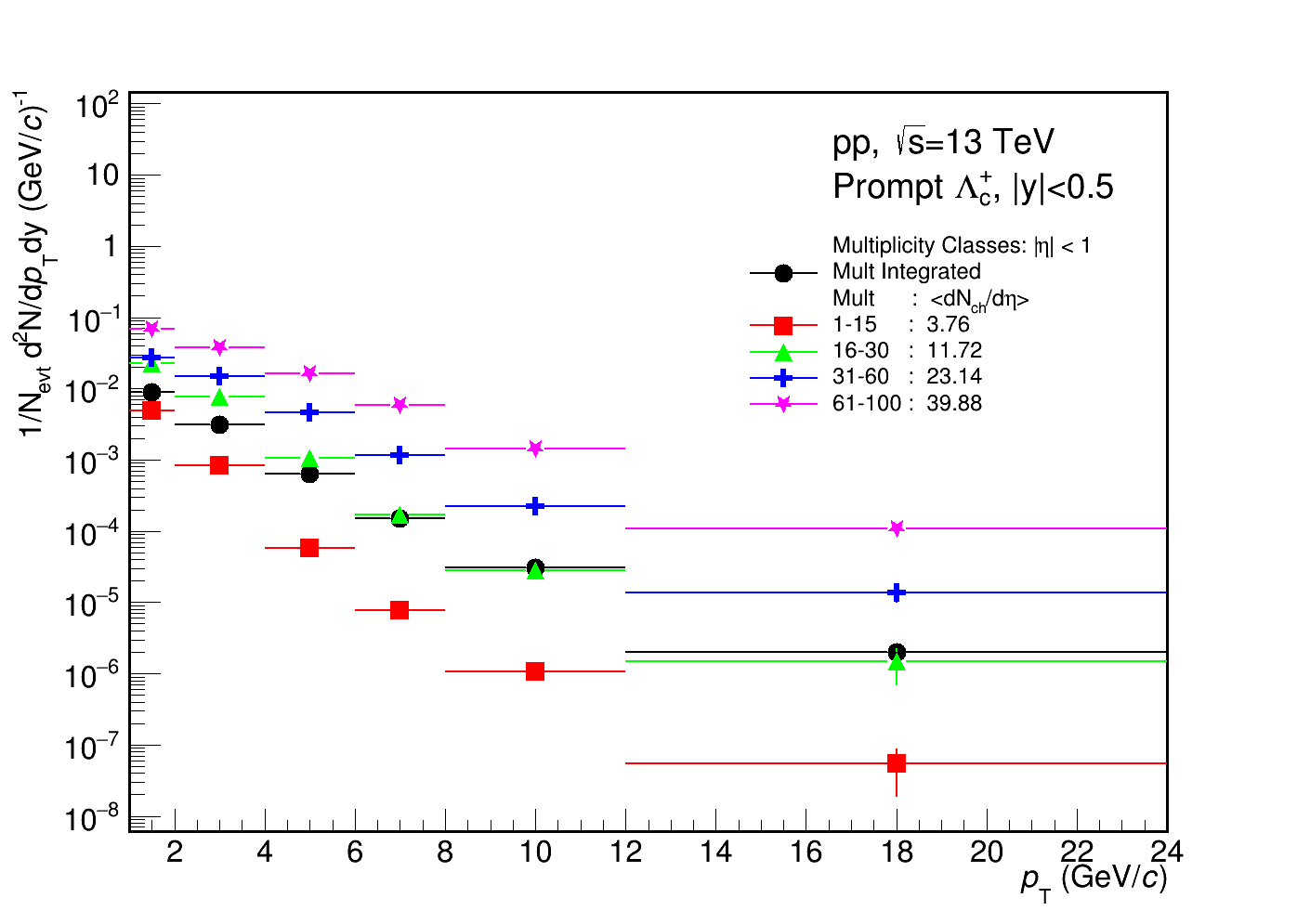}}
	\subfigure[] {\label{fig:b}\includegraphics[width=3.0in,height=2.0in]{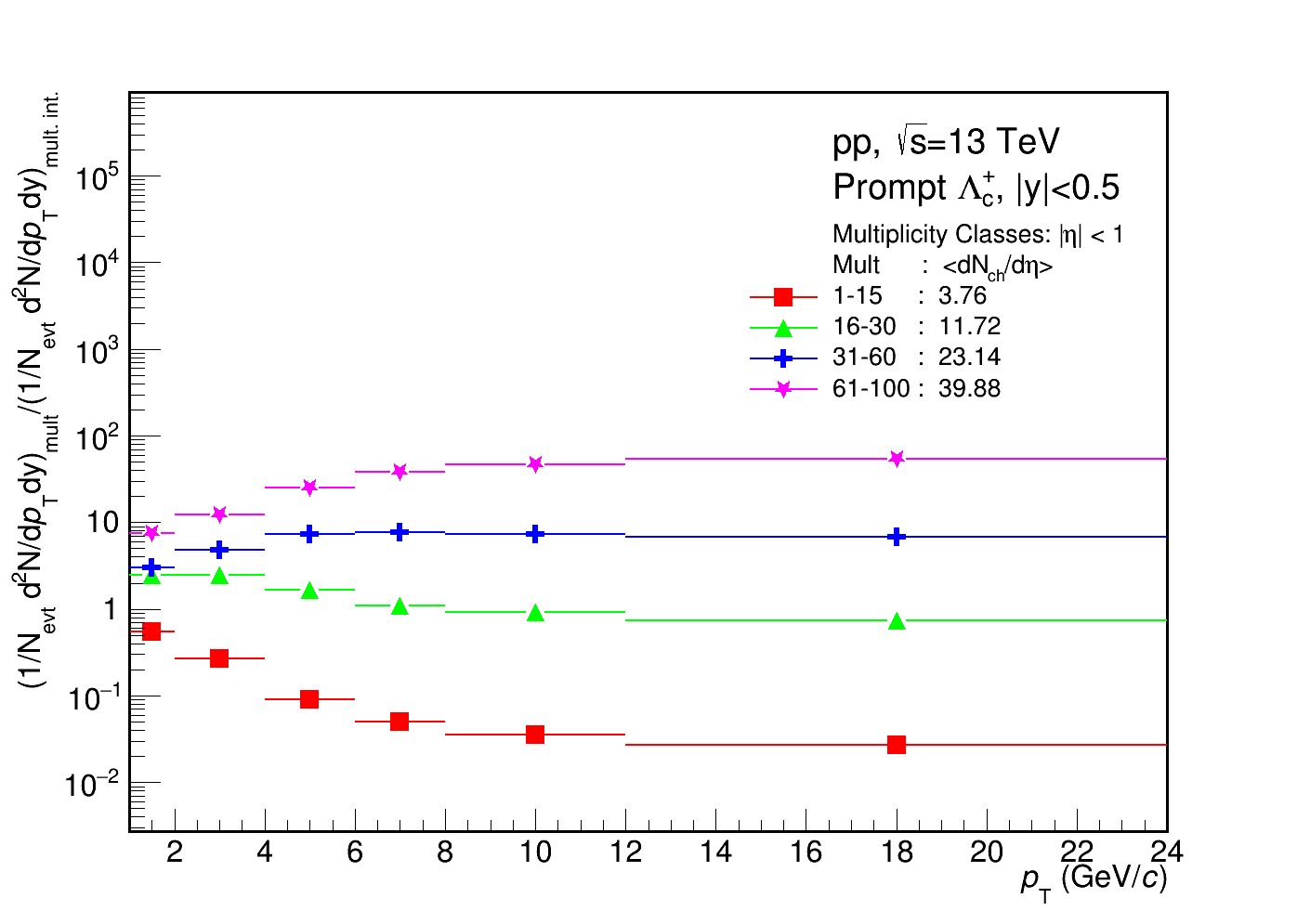}}
	\caption{(Left) Corrected spectra for $\Lambda^+_c$-baryon in the integrated and different multiplicity ranges. (Right) Ratio of corrected spectra in different multiplicity bins with respect to the spectra in the integrated multiplicity bin}
	\label{figureLambda}
\end{figure}

\textbf{A. Corrected p$_T$-spectra}\\
Fig.\ref{figureDzero} to .\ref{figureLambda} shows the transverse $p_T$ spectra for D$^0$, D$^+_s$ and $\Lambda_c^+$, respectively, on the left for all the multiplicity classes as well as the integrated multiplicity case [1,200]. A clear trend with multiplicity is seen for all studied hadrons. The spectra increases as we go from low multiplicity to high multiplicity region. To quantify this more, the ratios of the corrected yields in different multiplicity intervals with respect to the multiplicity integrated yield are also shown. From the ratios, it can be seen that besides the decrease or increase, which seem similar for the four studied hadrons, a softening or hardening in the $p_T$ spectra is observed as well. The softening of the spectra can be observed to be prominant in the low multiplicity events, whereas in case of high multiplicity events, hardening of the spectra is clearly visible. Thus, at high $p_T$, the charm-flavour hadron production is associated mostly with high multiplicity events.\\

\textbf{B. Strange to non-strange ratio}\\
\begin{figure}[h!]
	\centering     
	\subfigure[D$^+_s$/D$^0$ Ratio] {\label{fig:a}\includegraphics[width=3.0in,height=2.0in]{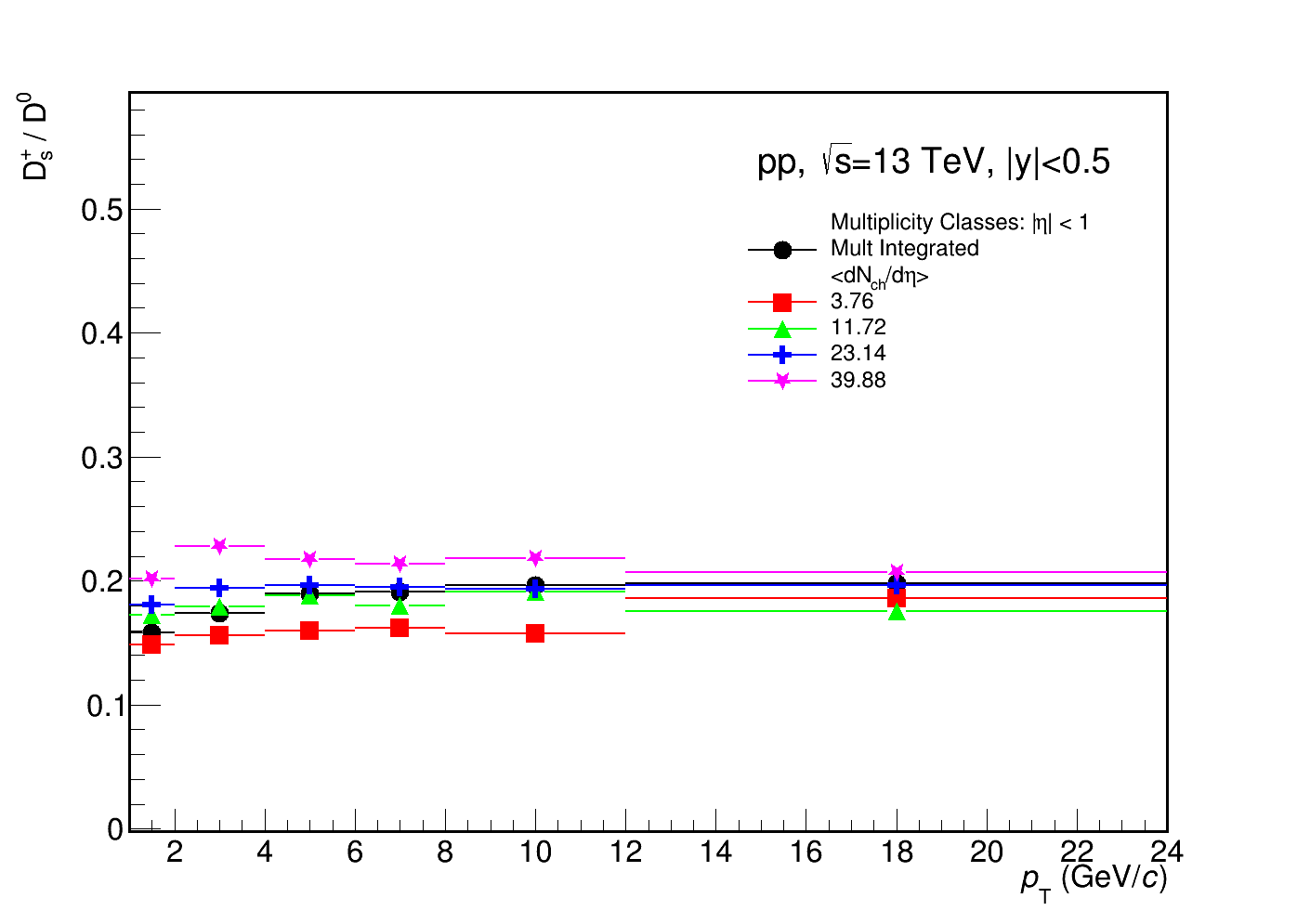}}
	\caption{Ratio of D$^+_s$ with respect to the D$^0$.}
	\label{ratios1}
\end{figure}
The charmed hadron ratios, especially D$^+_s$/D$^0$, are interesting to study the hadronisation mechanisms. In recent results from ALICE shows an enhancement in the integrated yields of strange and multi-strange particles relative to pions in the $pp$ events with  high charged-particle multiplicity. These measurements show a remarkable agreement with p-Pb collision results \cite{20,21}. The strangeness production in high-multiplicity events reaches values similar to those observed in Pb-Pb collisions, where a QGP is formed. In this study, the ratio of strange D-meson (D$_s^+$/D$^0$) show an enhancement with respect to the multiplicity. Fig.\ref{ratios1} shows the ratio of D$^+_s$  with respect to the D$^0$. The enhancement is not so strong at high $p_T$ but is somehow considerable at low $p_T$. \\

\textbf{B. Baryon to meson ratio}\\

\begin{figure}[h!]
	\centering     
	\subfigure[$\Lambda^+_c$/D$^0$ Ratio] {\label{fig:a}\includegraphics[width=3.0in,height=2.0in]{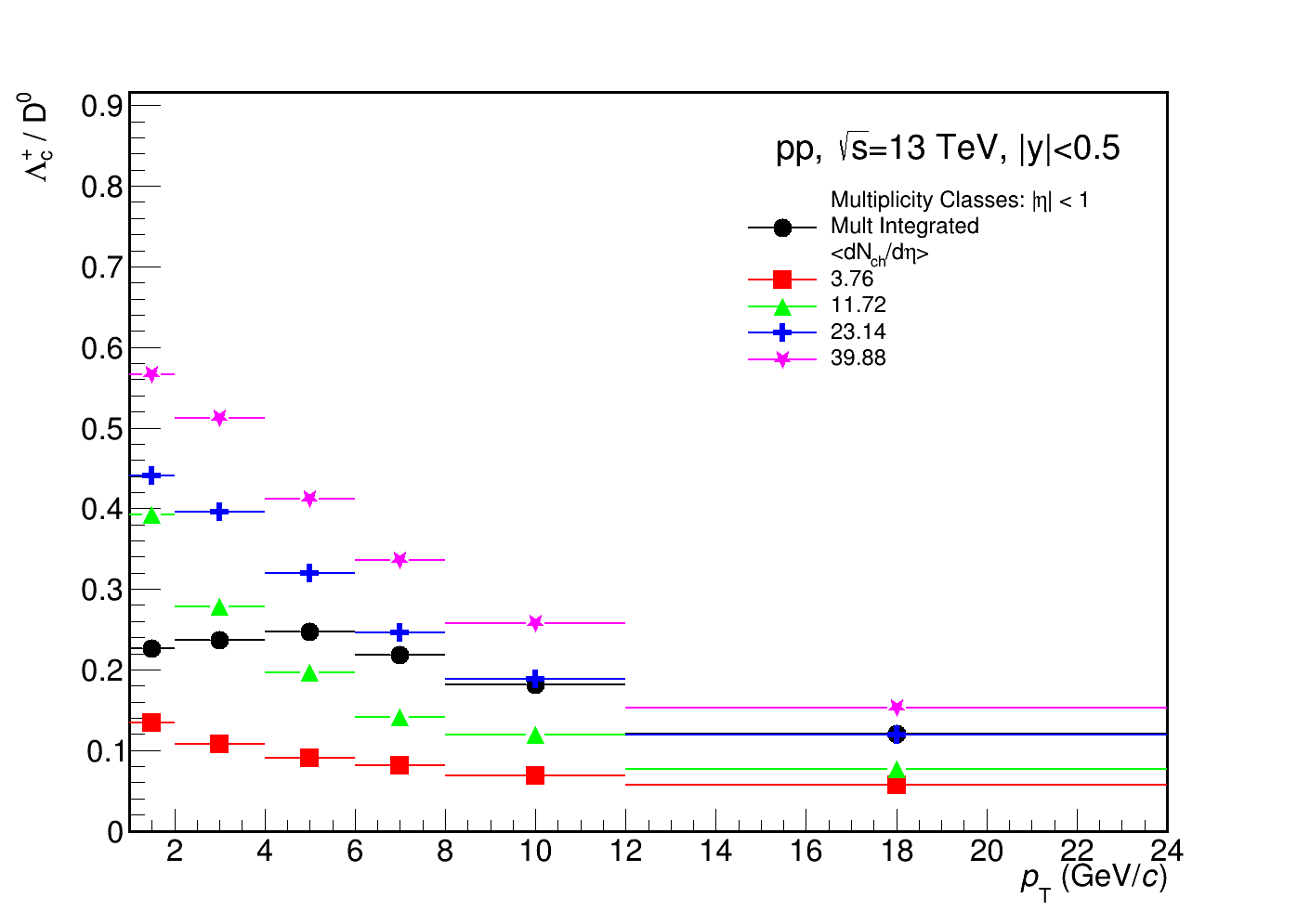}}
	\caption{Ratio of $\Lambda_c^+$ with respect to the D$^0$.}
	\label{ratios2}
\end{figure}
The $\Lambda^+_c$/D$^0$ ratios indicate a strong multiplicity dependence. In the recent results published by the ALICE and the CMS Collaboration, an enhancement of baryon-to-meson ratios ($p/\pi$ and $K/\pi$) at intermediate $p_T$ in high multiplicity pp and p–Pb collisions is observed. These results are similar to what is observed in heavy-ion collisions \cite{22,23}. Furthermore, it has been demonstrated that the evolution of the baryon-over-meson ratios as a function of $<dN_{ch}/d\eta>$ exhibits a universal pattern for all collision systems. This behavior might indicate a common mechanism at work that depends solely on final-state multiplicity density. The observed enhancement can be explained by coalescence model which considers hadronize-combination of constituent quarks \cite{24,25,26}. In this paper, we have also measured the baryon-to-meson ratio in different multiplicity regimes. The relative abundance of baryons and mesons can help to understand the process of fragmentation which is a dominating effect at high $p_T$. Due to MPIs, partons from jet in pp collisions can combine with quarks and antiquarks which are produced from MPIs to form hadrons. Since the momenta of partons from jets are significantly higher than the momenta of quarks and antiquarks produced from secondary MPIs, the resulting hadrons have momenta lower than independent fragmentation of jet partons. Fig.\ref{ratios2} shows the ratio of $\Lambda^+_c$ with respect to D$^0$. The $\Lambda^+_c$/D$^0$ ratios indicate a strong multiplicity dependence when compared to strange-to-nonstrange ratios (fig.\ref{ratios1}). The enhancement of $\Lambda^+_c$/D$^0$ ratio with multiplicity may indicate a contribution from jets in hard events resulting to the increase in the density of quarks and gluons. 

\section{Summary}
\label{section4}
In this contribution, the production of open charm hadrons viz. D$^0$, D$^+_s$ and $\Lambda^+_c$ has been studied in pp collisions at $\sqrt{s}$ = 13.6 TeV using CR2 tune in PYTHIA 8. The corrected $p_T$ spectra has been plotted for all the hadrons in the mid rapidity range $|y| <$ 0.5. The spectra has been calculated in four multiplicity ranges [1,15], [16,30], [31,60] and [61,100] and in the integrated multiplicity range [1,200]. The kinematic range of 1-24 GeV/\textit{c} is considered. The spectra of all the particles show a multiplicity dependence being smallest in the lowest multiplicity bin and highest in the largest multiplicity bin for all $p_T$ ranges. The results hint to the large production of open charm flavour in the high multiplicity pp events. The increased multiplicity of events may be attributed to hard MPIs which can induce a correlation between the yield of heavy quarks and the total charged particle multiplicity. In case of results from ALICE data in pp collisions at $\sqrt{s}$ = 13.6 TeV, the yield production of D$^+_s$ and $\Lambda_c^+$ and their ratio with D$^0$ have shown a similar trend with respect to the multiplicity~\cite{16b}.

\section{Acknowledgement}
We thankfully acknowledge all the authors of PYTHIA8 for their continuous support.

\begin{appendices}
	\section{Appendix}
\begin{figure}[h!]
		\centering     
		{\includegraphics[width=3.0in,height=2.0in]{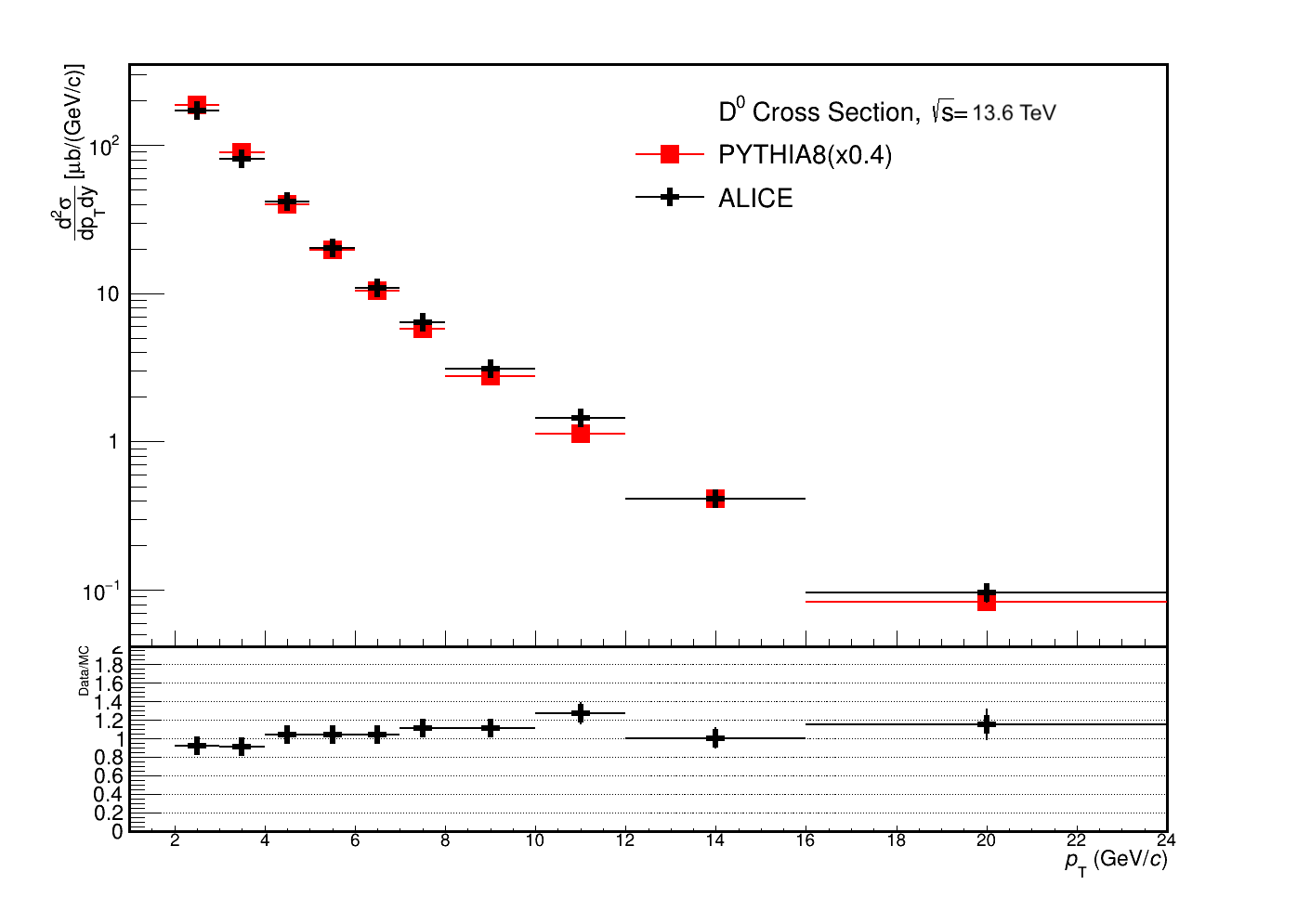}}  
		\caption{Figure show the comparison of D$^0$ cross-sections as a function of $p_T$ in pp collisions at $\sqrt{s}$ = 13.6 TeV as measured in PYTHIA 8 and ALICE data respectively. In the bottom panel the ratio of the cross-section as measured in ALICE is plotted with respect to the PYTHIA 8. The error bars are estimated using standard error propagation formula. }
		\label{crosssection}
	\end{figure}

	For checking the compatibility between PYTHIA 8 and the experimental data, we have compared the production cross-section of D$^0$ between experimental data from ALICE and PYTHIA 8 in the same kinematic range.  Fig.\ref{crosssection}  shows the comparison of D$^0$ production cross-section in pp collisions as a function of $p_T$ respectively. The black markers represent the data obtained from ALICE experiment \cite{27} and the red markers represents the results from PYTHIA 8 event generator in pp collisions at $\sqrt{s}$ = 13.6 TeV. In order to see how well the spectral shapes obtained from PYTHIA 8 simulation match with the experimental data, we have used some arbitrary multipliers. From the fig.\ref{crosssection}, it can be seen that PYTHIA 8 seems to reproduce similar spectral shapes as from experimental data for D$^0$.	
	
\end{appendices}
\end{document}